\documentclass{appolb}
\usepackage{epsfig}
\newcommand{\lesssim}{\raisebox{0.3mm}{\em $\, <$} 
\hspace{-3.3mm} \raisebox{-1.8mm}{\em $\sim \,$}}
\newcommand{\gtrsim}{\raisebox{0.3mm}{\em $\, >$}
\hspace{-3.3mm} \raisebox{-1.8mm}{\em $\sim \,$}}

\begin{document}
\title{
Search for sterile neutrinos at reactors with a small core
\thanks{
Presented at the XXXV International School of Theoretical
Physics ``Matter To The Deepest:
Recent Developments In Physics
of Fundamental Interactions'', Ustron, Poland,
September 12--18, 2011.
}%
}
\author{Osamu~Yasuda
\address{
Department of Physics, Tokyo Metropolitan University,
Hachioji, Tokyo 192-0397, Japan
}
}
\maketitle
\begin{abstract}
The sensitivity to the sterile neutrino mixing
at very short baseline reactor neutrino experiments
is investigated.  If the reactor core
is relatively large as in the case of 
commercial reactors,
then the sensitivity is lost for
$\Delta m^2 \gtrsim$ 1 eV$^2$ due to
smearing of the reactor core size.
If the reactor core is small
as in the case of the experimental fast
neutron reactor Joyo, the ILL
research reactor or the Osiris
reactor, on the other hand, then
sensitivity to $\sin^22\theta_{14}$
can be as good as 0.03 for
$\Delta m^2 \sim$ several eV$^2$
because of its small size.
\end{abstract}
  
\section{Introduction}
Schemes with sterile neutrinos have attracted a lot of
attention since the LSND group announced the
anomaly\,\cite{Athanassopoulos:1996jb,Athanassopoulos:1997pv,Aguilar:2001ty}
which would imply mass squared
difference of ${\cal O}$(1) eV$^2$
if it is interpreted as a phenomenon due to neutrino oscillation.
The standard three flavor scheme has only two
independent mass squared differences, i.e., $\Delta m^2_{21}=\Delta
m^2_\odot\simeq 8\times 10^{-5}$eV$^2$ for the solar neutrino
oscillation, and
$|\Delta m^2_{31}|=\Delta m^2_{\mbox{\scriptsize\rm atm}}\simeq
2.4\times 10^{-3}$eV$^2$ for the atmospheric neutrino
oscillation.
To accommodate a neutrino oscillation scheme to the LSND
anomaly, therefore, the extra state should be
introduced.  This extra state should be sterile neutrino, which is
singlet with respect to the gauge group of the Standard Model,
because the number of weakly interacting light neutrinos
should be three from the LEP data\,\cite{Nakamura:2010zzi}.

Recently sterile neutrino scenarios are becoming popular again
because of a few reasons.  One is the data
of the MiniBooNE experiment which been performed to test
the LSND anomaly.  Although their data on the neutrino
mode\,\cite{AguilarArevalo:2007it} disfavors the region
suggested by LSND, their data on the anti-neutrino
mode\,\cite{AguilarArevalo:2010wv} seems to be consistent
with that of LSND.  The second one is the so-called
reactor anomaly.  The flux of the reactor neutrino was
recalculated in Ref.\,\cite{Mueller:2011nm} recently and
it was claimed that the normalization is shifted by about
+3\% on average.  This claimed is qualitatively consistent
with an independent calculation in
Ref.\,\cite{Huber:2011wv}.  If their claim on the reactor
neutrino flux is correct, then neutrino oscillation with
$\Delta m^2\gtrsim$1eV$^2$ may be concluded from a
re-analysis of 19 reactor neutrino results at short
baselines\,\cite{Mention:2011rk}.  The third one is the
so-called gallium anomaly.
The data of the gallium solar neutrino calibration experiments
indicates deficit of $\nu_e$ and it may imply
neutrino oscillation\,\cite{Giunti:2010wz}.

It has been known that reactor experiments with more than
one detector has a possibility to measure $\theta_{13}$
precisely because some of the systematic errors can be
canceled by the near-far detector
complex\,\cite{Kozlov:2001jv,Minakata:2002jv,Huber:2003pm,Anderson:2004pk}.
Three experiments\,\cite{Ardellier:2004ui,Ahn:2010vy,Guo:2007ug}
are now either running or expected to start soon
to measure $\theta_{13}$.
In the standard three flavor case with
$|\Delta m^2_{31}|=2.4\times10^{-3}$eV$^2$,
it was shown assuming infinite statistics
that the optimized baseline lengths $L_F$ and $L_N$ for the far
and near detectors are $L_F\simeq$1.8km and $L_N\simeq$0km in the rate
analysis\,\cite{Yasuda:2004dd,Sugiyama:2004bv}, while they are
$L_F\simeq$10.6km and $L_N\simeq$8.4km in the spectrum
analysis\,\cite{Sugiyama:2005ir}.
To justify the assumption
on negligible statistical errors for $L\sim$10km, unfortunately,
one would need unrealistically huge detectors,
so one is forced to choose the baseline
lengths which are optimized for the rate analysis
for $\Delta m^2=2.4\times10^{-3}$eV$^2$.
On the other hand, if one performs an oscillation
experiment to probe $\Delta m^2\sim{\cal O}$(1) eV$^2$,
it becomes realistic to place the detectors at the baseline lengths
which are optimized for the spectrum analysis~(See Sect.
4 in the published version of Ref.\,\cite{Sugiyama:2005ir}).

In this talk I would like to discuss the sensitivity of very short
line reactor experiments to the sterile neutrino mixing
for $\Delta m^2\sim{\cal O}$(1) eV$^2$ in the so-called
(3+1)-scheme\,\cite{Yasuda:2011np}.
Proposals have been made to test the
bound of the Bugey reactor
experiment\,\cite{Declais:1994su} on the sterile neutrino
mixing angle using a reactor\,\cite{nucifer,panda}\footnote{
See, e.g., Refs.\,\cite{Sugiyama:2005ir} (the published
version), \cite{Latimer:2007qe,deGouvea:2008qk} for
earlier works on search for sterile neutrinos at a
reactor.}, an
accelerator\,\cite{Agarwalla:2009em,Baibussinov:2009tx}, and a
$\beta$-source\,\cite{Cribier:2011fv,Dwyer:2011xs}.

\section{Four neutrino schemes}
Four-neutrino schemes consist of one extra sterile state
and the three weakly interacting ones.
The schemes are
called (3+1)- and (2+2)-schemes,
depending on whether one or two
mass eigenstate(s) are separated from the others by the largest
mass-squared difference $\sim{\cal O}$(1) eV$^2$. 
The (2+2) scheme is excluded by the solar and
atmospheric neutrino data\,\cite{Maltoni:2004ei}, so
I will not discuss the (2+2) schemes here.
In the (3+1) scheme,
the phenomenology of solar and atmospheric oscillations is
approximately the same as that of the three flavor framework,
so there is no tension
between the solar and atmospheric constraints.
However, the (3+1) scheme
has a problem in accounting for LSND and all other negative results
of the short baseline experiments.
To explain the LSND data while satisfying the
constraints from other disappearance experiments, the oscillation
probabilities of the appearance and disappearance channels should
satisfy the following relation \cite{Okada:1996kw,Bilenky:1996rw}:
\begin{eqnarray}
\sin^22\theta_{\mbox{\rm\tiny LSND}}(\Delta m^2)
<\frac{1}{4}\,\sin^22\theta_{\mbox{\rm\scriptsize Bugey}}(\Delta m^2)
\cdot
\sin^22\theta_{\mbox{\rm\tiny CDHSW}}(\Delta m^2)
\label{relation31}
\end{eqnarray}
where $\theta_{\mbox{\rm\tiny LSND}}(\Delta m^2)$,
$\theta_{\mbox{\rm\tiny CDHSW}}(\Delta m^2)$,
$\theta_{\mbox{\rm\scriptsize Bugey}}(\Delta m^2)$ are the value of
the effective two-flavor mixing angle as a function of the mass
squared difference $\Delta m^2$ in the allowed region for LSND
($\bar{\nu}_\mu\rightarrow\bar{\nu}_e$), the CDHSW experiment
\cite{Dydak:1983zq} ($\nu_\mu\rightarrow\nu_\mu$), and the Bugey
experiment \cite{Declais:1994su}
($\bar{\nu}_e\rightarrow\bar{\nu}_e$), respectively.  The reason that
the (3+1)-scheme to explain LSND has been disfavored is because
Eq.~(\ref{relation31}) is not satisfied for any value of $\Delta m^2$,
if one adopts the allowed regions in Refs.\,\cite{Dydak:1983zq} and
\cite{Declais:1994su}.
If the flux of the reactor neutrino is slightly
larger than the one used in the Bugey
analysis\,\cite{Declais:1994su}, however, the allowed region
becomes slightly wider and one has more chance to satisfy
Eq.~(\ref{relation31})\footnote{ 
Although the situation of the (3+1)-scheme is improved slightly after
Refs.\,\cite{Mueller:2011nm,Mention:2011rk},
the improvement
is not sufficient enough to have a satisfactory fit to all
the data, according to Ref.\,\cite{schwetz}.}.

I will use the following parametrization for the mixing
matrix\,\cite{Maltoni:2007zf}:
\begin{eqnarray}
    U =
    R_{34}(\theta_{34} ,\, 0) \; R_{24}(\theta_{24} ,\, 0) \;
    R_{23}(\theta_{23} ,\, \delta_3) \;
    R_{14}(\theta_{14} ,\, 0) \; R_{13}(\theta_{13} ,\, \delta_2) \; 
    R_{12}(\theta_{12} ,\, \delta_1) \,,
    \nonumber
\end{eqnarray}
where $R_{jk}(\theta_{jk},\ \delta_l)$ are the complex
rotation matrices in the $jk$-plane defined as
\begin{eqnarray}
&{\ }&[R_{jk}(\theta_{jk},\ \delta_{l})]_{pq}
\nonumber\\
& = &\delta_{pq}+(\cos \theta_{jk}-1)
(\delta_{jp}\delta_{jq}+\delta_{kp}\delta_{kq})
+\sin \theta_{jk}
(e^{-i\delta_l}\delta_{jp}\delta_{kq}
-e^{i\delta_l}\delta_{jq}\delta_{kp}).
\nonumber
\end{eqnarray}
With this parametrization, for the
very short baseline reactor experiments,
where the average neutrino energy $E$ is
approximately 4MeV and the baseline length is
about 10m, I have $|\Delta m^2_{jk}L/4E|\ll 1~(j,k=1,2,3)$,
so that the disappearance probability
is given by 
\begin{eqnarray}
    P(\bar{\nu}_e\to\bar{\nu}_e) =
    1-\sin^22\theta_{14}\,\sin^2
\left(\frac{\Delta m^2_{41}L}{4E}
\right)
    \label{prob}
\end{eqnarray}
to a good approximation.
So the analysis of the (3+1)-scheme is reduced to that of a two
flavor framework with the oscillation
parameters ($\Delta m^2_{41}$, $\sin^22\theta_{14}$).

\section{Sensitivity to $\sin^2{2\theta_{14}}$
by a spectral analysis}

Throughout my talk I discuss the case with
a single reactor and two detectors.
I assume here that
the near and far detectors are identical and they have the
same sizes of systematic errors.
The conditions of the detectors are assumed to be
the same as those of the Bugey experiment,
i.e., liquid scintillation detector of volume 600 liters
with the detection efficiency which yields
about 90,000 events at $L$=15m from a reactor of
a power 2.8GW after running for 1800 hours.

To evaluate the sensitivity to $\sin^2{2\theta_{14}}$,
let us introduce the following
$\chi^2$ which was adopted in Ref.\,\cite{Sugiyama:2005ir}
~(See Ref.\,\cite{Yasuda:2011np} for details):
\begin{eqnarray}
\hspace*{-20mm}
\displaystyle
\chi^2&=&\min_{\alpha's}\Bigg\{
\displaystyle\sum_{A=N,F}\sum_{i=1}^n
\frac{1} {(t^A_i\sigma^A_i)^2}
\left[m^A_i-t^A_i(1+\alpha+\alpha^A+\alpha_i)
-\alpha_{\mbox{\scriptsize\rm cal}}^A t^A_iv^A_i
\right]^2\nonumber\\
&+&\displaystyle\sum_{A=N,F}\left[\left(
\frac{\alpha^A} {\sigma_{\mbox{\scriptsize\rm dB}}}\right)^2
+\left(\frac{\alpha_{\mbox{\scriptsize\rm cal}}^A} {\sigma_{\mbox{\scriptsize\rm cal}}}\right)^2\right]
+\displaystyle\sum_{i=1}^n\left(
\frac{\alpha_i} {\sigma_{\mbox{\scriptsize\rm Db}}}\right)^2
+\left(\frac{\alpha} {\sigma_{\mbox{\scriptsize\rm DB}}}\right)^2\Bigg\}.
\label{chipull}
\end{eqnarray}
$\chi^2$ stands for a quantity which expresses
how much deviation we have between
the numbers of events with and without
oscillations, compared with the
experimental errors.
In Eq.(\ref{chipull}), $m^A_i$ is the number of events to be measured
at the near ($A=N$) and far ($A=F$)
for the $i$-th energy bin
with the neutrino oscillation,
and $t^A_i$ is the theoretical prediction without the oscillation.
$(\sigma^A_i)^2$ is the uncorrelated error
which consists of the statistical plus uncorrelated bin-to-bin
systematic error:
$(t^A_i\sigma^A_i)^2
=t^A_i+\left(t^A_i\sigma^A_{\mbox{\scriptsize\rm db}}\right)^2$,
where $\sigma^A_{\mbox{\scriptsize\rm db}}$ is the uncorrelated bin-to-bin
systematic error.
$\alpha^A~(A=N,F)$ is a variable
which introduces the detector-specific
uncertainties $\sigma_{\mbox{\scriptsize\rm dB}}$ of the
near and far detectors.
$\alpha_i~(i=1,\cdots,n)$ is a variable for
an uncertainty $\sigma_{\mbox{\scriptsize\rm Db}}$ of the
theoretical prediction for each energy bin which
is uncorrelated between different energy bins.\footnote{
The first suffix of $\sigma$ stands for the property for
the systematic error with respect to the detectors while
the second is with respect to bins, and capital (small)
letter stands for a correlated (uncorrelated) systematic
error.}
$\alpha_{\mbox{\scriptsize\rm cal}}^A~(A=N,F)$ is a variable which introduces 
an energy calibration uncertainty $\sigma_{\mbox{\scriptsize\rm cal}}$
and comes in the theoretical prediction in the form
of $(1+\alpha_{\mbox{\scriptsize\rm cal}}^A)E$ instead of the observed energy $E$.
$v^A_i$ is the deviation divided by
the expected number of events from the
theoretical prediction $t^A_i$ due to
the energy calibration uncertainty.
Here I take the following reference values for the systematic errors:
$\sigma_{\mbox{\scriptsize\rm db}}=0.5\%$,
$\sigma_{\mbox{\scriptsize\rm dB}}=0.5\%$,
$\sigma_{\mbox{\scriptsize\rm Db}}=2\%$,
$\sigma_{\mbox{\scriptsize\rm DB}}=3\%$,
$\sigma_{\mbox{\scriptsize\rm cal}}=0.6\%$.

\subsection{Commercial reactors}
First of all, I will consider
a commercial reactor whose
thermal power is 2.8GW
and I will assume that the dimension of
its core is 4m in diameter and 4m in height.

\begin{figure}[h]
\begin{center}
\epsfig{file=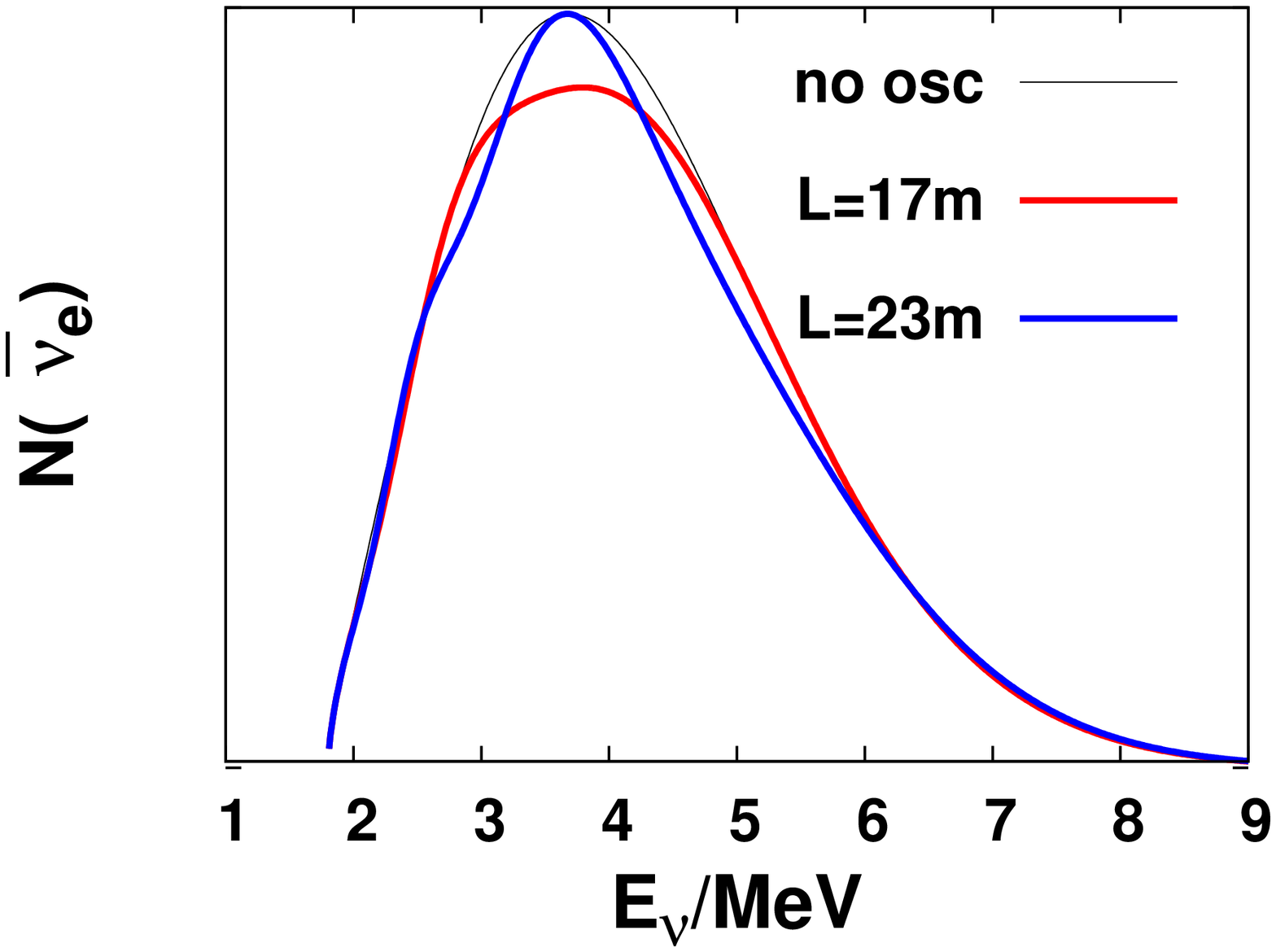,width=9.8cm}
\epsfig{file=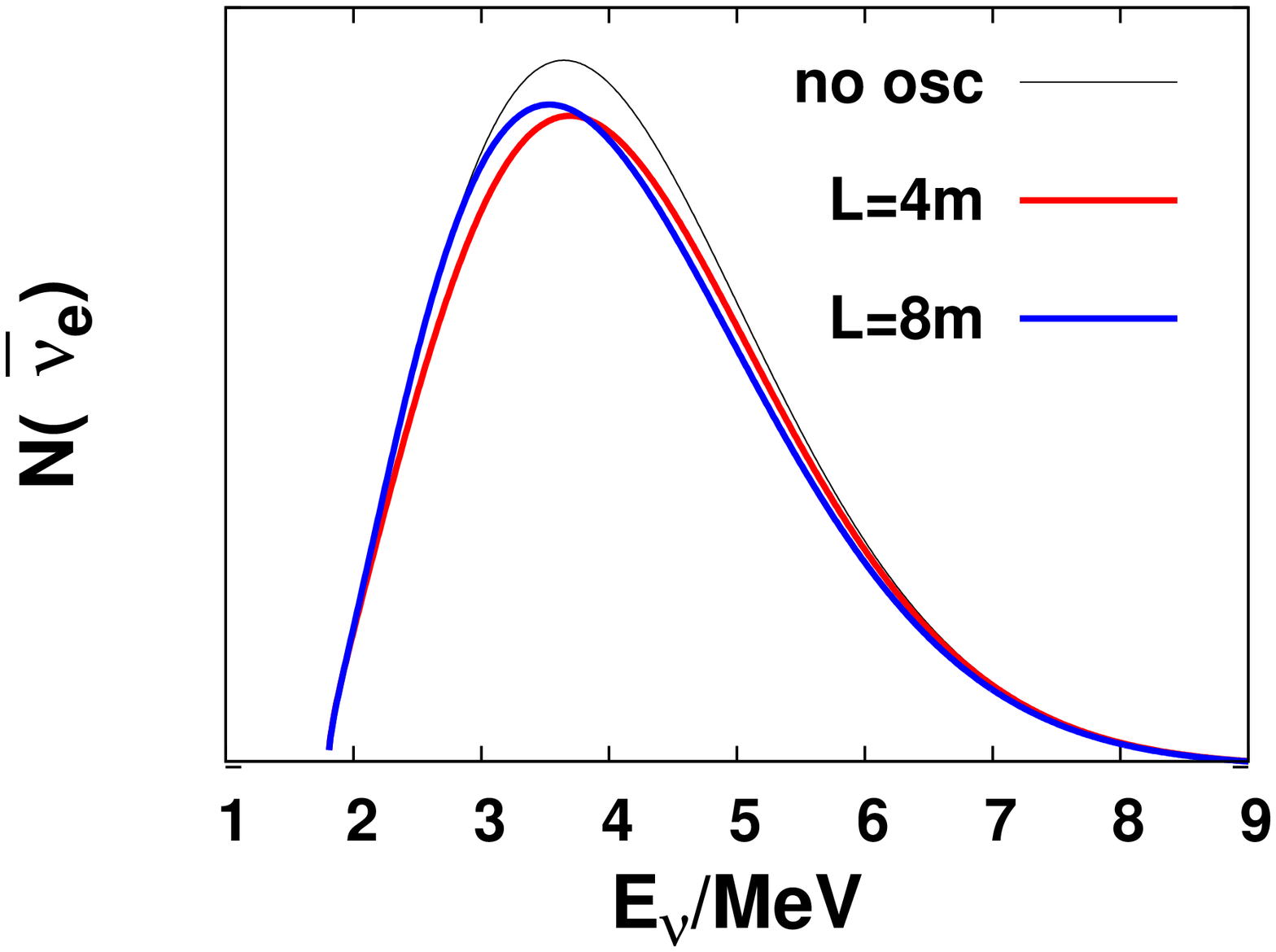,width=9.8cm}
\end{center}
\vspace*{-5mm}
\caption{
The energy spectrum with
neutrino oscillations at the two different
detectors and the one
without oscillations.  The
optimized baseline lengths give
maximum difference in the
distortions in the energy
spectrum.  The upper panel:
the case of a commercial
reactor.  The lower panel:
the case of a research reactor.
}
\label{fig1}
\end{figure}

$\chi^2$ in Eq.~(\ref{chipull}) is computed
numerically in the case of  $\Delta m^2_{41}=1$eV$^2$
as a function of the baseline lengths $L_N$ and $L_F$
of the two detectors, and
the baseline lengths $L_N$ and $L_F$ are varied
to optimize the sensitivity to $\sin^22\theta_{14}$.
It is found that the set $(L_N, L_F)\simeq$ (17m, 23m)
gives the optimum.  In contrast to the
rate analysis, in which the optimized
baseline length of the near detector
is $L_N$=0m to avoid oscillations,
the spectrum analysis with
$(L_N, L_F)=$ (17m, 23m) looks
at the difference between the
maximum and minimum of the spectrum shape
with neutrino oscillations
at $L_N$ and $L_F$
mainly for the energy region $E_\nu\sim$ 4MeV
where the number of events are expected to be
the largest (See the upper panel in Fig.\,\ref{fig1}).
Unlike the case of infinite statistics\,\cite{Sugiyama:2005ir},
the statistical errors are important in the
present setup of the detectors, and
longer baseline lengths are disfavored.

\begin{figure}[h]
\vspace*{-7mm}
\epsfig{file=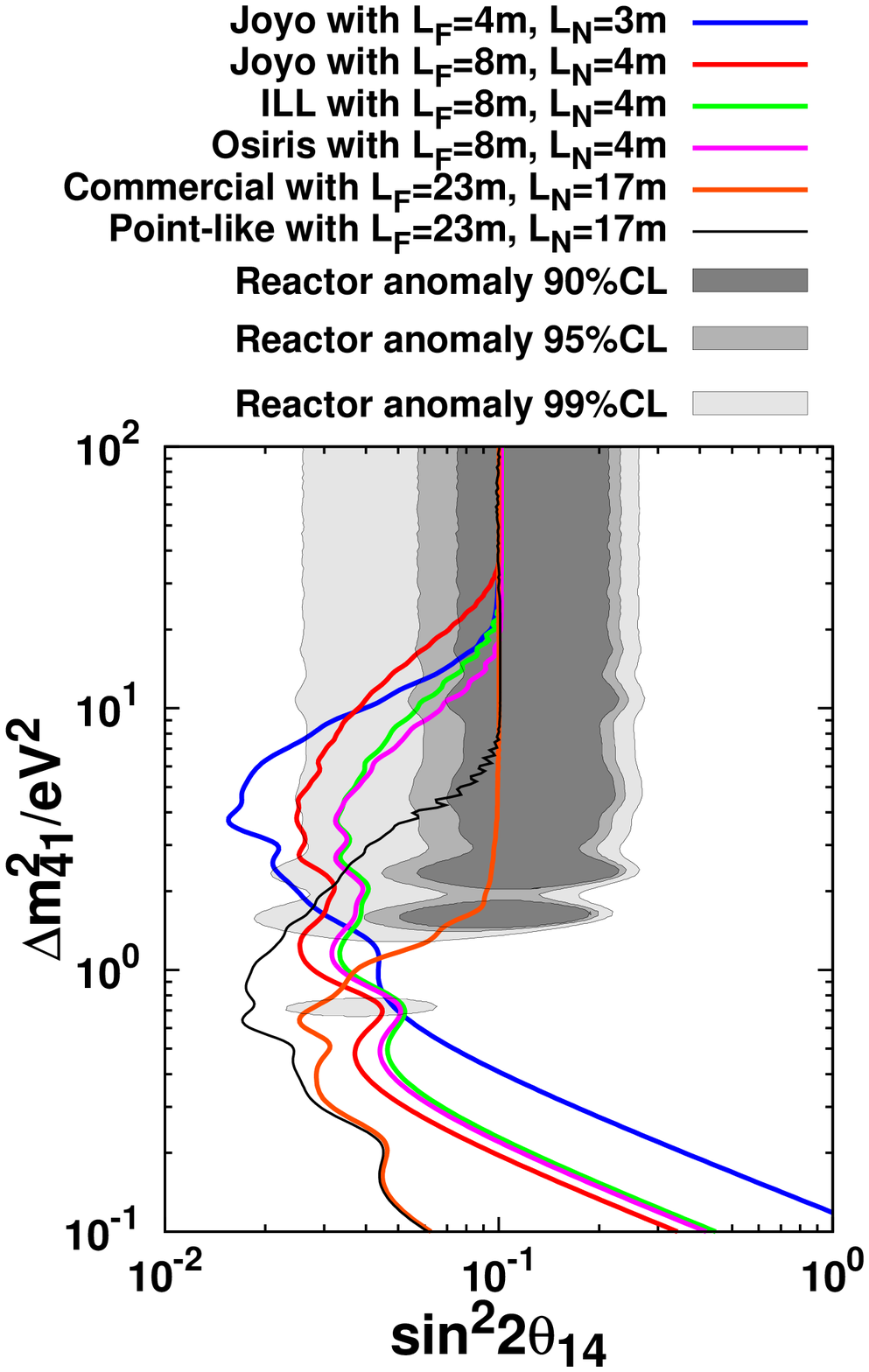,width=10.7cm}
\vspace*{-4mm}
\caption{
The sensitivity to $\sin^22\theta_{14}$
of each reactor with the two detectors at its
optimum baseline lengths.
Also shown as a shaded area
is the region given in Ref.\,\cite{Mention:2011rk}
from the combination of the reactor neutrino experiments, Gallex and
Sage calibration sources experiments, the MiniBooNE reanalysis of 
Ref.\,\cite{Giunti:2010wz}, and the ILL-energy spectrum distortion.}
\label{fig2}
\end{figure}

The sensitivity
to $\sin^2{2\theta_{14}}$ in the case of the baseline
lengths $(L_N, L_F)=$ (17m, 23m) is shown in
Fig.\,\ref{fig2} as a function of
$\Delta m^2_{41}$ (the line referred to as
``Commercial'').
The region suggested by combination of
the reactor and gallium anomalies and
the MiniBooNE data
is also given in Fig.\,\ref{fig2} for comparison.
For $\Delta m^2_{41}\gtrsim 2$eV$^2$,
the sensitivity is
no better than 0.1, which is basically
the result of the rate analysis.
The sensitivity in the case of a hypothetical
point-like reactor, where all the conditions
for the detectors are the same, is also given in
Fig.\,\ref{fig2} for comparison (the line referred to as
``Point-like'').
Fig.\,\ref{fig2} indicates that the sensitivity
would be as good as several $\times 10^{-2}$
for a few eV$^2$, if the core were point-like.
So we can conclude that we have poor sensitivity
for $\Delta m^2_{41}\gtrsim$ 2eV$^2$ because of
the smearing effect of the
finite core size of the reactor.

\subsection{Research reactors}
In the previous subsection, we have seen that
the sensitivity to $\sin^2{2\theta_{14}}$ is
lost because of the smearing effect of finite
core size.
Next, I would like to discuss three
research reactors, 
Joyo\,\cite{joyo} with MK-III
upgrade\,\cite{joyomk3}, the ILL
research reactor\,\cite{ill}, and the Osiris
research reactor\,\cite{osiris}.
They all have a relatively small size and
a relatively large thermal power.

Joyo is an experimental fast breeder reactor
and the dimension of its core is
0.8m in diameter and 0.5m in height,
and its thermal power is 140MW.
The ILL (Osiris) research reactor is
a thermal neutron reactor with high
enrichment uranium ${}^{235}$U, and the dimension of its
core is 0.4m in diameter and 0.8m in height
(0.57m$\times$0.57m$\times$0.6m) and its
thermal power is 58MW (70MW), respectively.

Again $\chi^2$ in Eq.~(\ref{chipull}) is computed 
numerically in each case, and it is optimized
with respect to $L_N$ and $L_F$.
The optimum set of the baseline lengths turns out
to be $(L_N, L_F)\simeq$ (4m, 8m) for
$\Delta m^2_{41}=1$eV$^2$ for all the three cases.
The lower panel in Fig.\,\ref{fig1} shows
the spectrum distortion in the case of $L$=4m, 8m.

The sensitivity to $\sin^2{2\theta_{14}}$ is shown
in Fig.\,\ref{fig2} as a function of
$\Delta m^2_{41}$ in the case of the sets
of the baseline lengths $(L_N, L_F)=$ (4m, 8m)
for the three cases
and $(L_N, L_F)=$ (3m, 4m) for Joyo.
From Fig.\,\ref{fig2} it is clear that
the sensitivity of an experiment
with a small core reactor is better that
that with a commercial reactor for
2eV$^2$$\lesssim\Delta m^2_{41}\lesssim10$eV$^2$.

\section{Discussion and Conclusion}

In the framework of the (3+1)-scheme,
the sensitivity to $\sin^22\theta_{14}$
of very short baseline reactor oscillation experiments
was studied by a spectrum analysis.
The assumptions are that one has two detectors
whose size and efficiency are exactly the same as
those used at the Bugey experiment
and $\chi^2$ is optimized with respect to the
positions of the two detectors.

In the case of a commercial reactor,
by putting the detectors at $L_N=$ 17m and
$L_F=$ 23m, one obtains the sensitivity
as good as several $\times 10^{-2}$ for
$\Delta m^2_{41} \lesssim 1$eV$^2$, but
the sensitivity is lost above 1eV$^2$
due to the smearing of the finite core size.

In the case of a research reactor
with a small core (such as Joyo, ILL, Osiris),
on the other hand, one obtains the sensitivity
as good as a several $\times 10^{-2}$ for
1eV$^2$$\lesssim\Delta m^2_{41} \lesssim 10$eV$^2$
if the detectors are located at $L_N=$ 4m and
$L_F=$ 8m.

In all the cases discussed above with the Bugey-like detector
setup, the statistical errors are dominant.  The reason
that the case of the research reactors (Joyo, ILL,
Osiris) is competitive despite its small power is because
the total numbers of events at $L\sim$ several meters are
comparable to those of the case with a commercial reactor
at $L\sim$ a few $\times$ 10 meters.

To turn this idea into reality, there are two experimental
challenges.  One is to put detectors at a location very
near to a research reactor.  The other one is to
avoid potentially huge backgrounds from the
reactor.\footnote{
An experiment\,\cite{Furuta:2011iu} was performed
to detect neutrinos from a fast neutron reactor
at Joyo, but unfortunately they did not get
sufficient statistical significance.}

Nevertheless, since the best fit point
($\Delta m^2_{41}$, $\sin^22\theta_{14}$) $\sim$
(2eV$^2$, 0.1) obtained in Ref.\,\cite{Mention:2011rk}
lies within the excluded region in Fig.\,\ref{fig2},
the experiment at these research reactors
offers a promising possibility.

\section*{Acknowledgments}
The author would like to thank Marek Biesiada and
other organizers for invitation and hospitality
during the conference.
He would also like to thank F.~Suekane and G.~Mention
for useful correspondence and K.~Schreckenbach
for discussions on research reactors.
This work was partly supported by Grants-in-Aid for Scientific Research
of the Ministry of Education, Culture, Sports, Science, and Technology,
under Grant No.\ 21540274.


\end{document}